\documentclass[final,3p,times]{elsarticle}

\usepackage{amssymb}
\usepackage{amsmath}
\usepackage{slashed}
\usepackage{xcolor}
\usepackage{threeparttable}
\usepackage{subcaption}

\begin{document}

\begin{frontmatter}



\title{Nuclear matter properties and neutron star structures from an extended linear sigma model}


\author{Yao Ma} 
\ead{mayao@nju.edu.cn}

\affiliation{organization={School of Frontier Sciences, Nanjing University},
            addressline={1520 Taihu Avenue}, 
            city={Suzhou},
            postcode={215123}, 
            country={China}}

\begin{abstract}
The properties of nuclear matter and the structures of neutron stars are analyzed with a baryonic extended linear sigma model in mean-field approximation, where the masses of baryons and mesons are generated via the spontaneous chiral symmetry breaking.
The couplings between the iso-scalar scalar meson and nucleons, $g_{\sigma NN}$, the iso-vector scalar meson and nucleons, $g_{a_0 NN}$, and the four-vector meson couplings play an important role in the properties of nuclear matter and neutron stars.
The introduction of the $\delta$ meson leads to a plateau structure of the symmetry energy, $E_{\rm sym}(n)$, at intermediate densities, which is crucial to the consistency of neutron skin thickness of $^{208}$Pb and the tidal deformability of a canonical neutron star.
The explicit chiral symmetry breaking term is then introduced with a constant background field, $\xi$, which can be related to the current quark mass and thus the pion-nucleon sigma term, $\sigma_{\pi N}$.
A negative $\sigma_{\pi N}$ leads to a stiffer EOS of neutron star matter and thus a larger maximum mass of neutron stars, but the value of $\sigma_{\pi N}$ needed to satisfy the astrophysical constraints is negative, not positive as the vacuum value.
The study may provide insights into the running behaviors of the parameters in the low-energy effective model to give the density-dependent description for the EOS of neutron star matter.
\end{abstract}




\end{frontmatter}



\section{Introduction}
\label{sec:intro}
Nuclear matter (NM) properties and its equation of state (EOS) are crucial for understanding the low-energy strong interaction at densities, but they are still not well understood due to the non-perturbative nature of QCD~\cite{Fukushima:2010bq,Lattimer:2015nhk,Chen:2017och,Baym:2017whm,Gil:2018yah,Ma:2019ery,Li:2019xxz,Sorensen:2023zkk}.
With the development of astrophysical observations, especially the detection of gravitational waves from neutron star (NS) mergers~\cite{Abbott:2017oio}, more constraints can be placed on the EOS of dense matter and the structure of NSs, which offers a good window to understand the properties of NM and the low-energy strong interaction at densities.
However, the parametrization of strong interaction at low energies in the past studies, focusing on the properties of NM and structures of NSs, is usually phenomenological, and the connection to QCD is not clear, e.g., one-boson-exchange models~\cite{Holinde:1975vg, Erkelenz:1974uj, Nagels:1977ze, Machleidt:1989tm} and relativistic mean-filed (RMF) models~\cite{Walecka:1974qa,Serot:1984ey,Serot:1997xg}.

Considering the importance of iso-scalar scalar meson \(\sigma\) and iso-vector scalar meson \(\delta\) in determining the properties of NM and its EOS~\cite{Holinde:1975vg, Erkelenz:1974uj, Nagels:1977ze, Machleidt:1989tm,Kubis:1997ew,Hofmann:2000vz,Liu:2001iz,Chen:2007ih,Roca-Maza:2011alv,Wang:2014jmr,Typel:2020ozc,Kong:2025dwl}, we constructed a baryonic extended linear sigma model (bELSM) in Refs.~\cite{Ma:2023eoz,Ma:2025llw} based on the extended linear sigma model~\cite{Fariborz:2005gm,Fariborz:2007ai,Fariborz:2009cq}.
In this framework, the \(\sigma\) and \(\delta\) mesons are the mixing states of 2-quark and 4-quark configurations to cope with the P-wave problem of the lightest scalar meson~\cite{ParticleDataGroup:2024vhw}, and baryons are introduced with diquark approximation~\cite{Olbrich:2015gln}.
The spectra of mesons and baryons at vacuum can be obtained via the spontaneous chiral symmetry breaking, which presents the symmetry pattern of the low-energy strong interaction.
By fitting the spectra of mesons and baryons at vacuum and the properties of NM around saturation density, the parameters of the model can be fixed, then the structure of NSs can be obtained by solving the Tolman-Oppenheimer-Volkoff (TOV) equations with the EOS of NM under RMF approximation~\cite{Tolman:1939jz,Oppenheimer:1939ne}.

It's found that a plateau structure of symmetry energy, \(E_{sym}(n)\), appears at intermediate density regions, \(\sim 2n_0\), which is crucial to the consistency of neutron skin thickness of \(^{208}\)Pb~\cite{PREX:2021umo,Reed:2021nqk} and the tidal deformability of a canonical NS, \(\Lambda_{1.4}\)~\cite{Abbott:2017oio}.
This aligns with the previous studies in Refs.~\cite{Paeng:2017qvp,Ma:2018xjw,Ma:2018jze,Ma:2021nuf,Lee:2021hrw,Zabari:2018tjk,Miyatsu:2022wuy,Li:2022okx}.
Besides, we also find that the structure of NSs is sensitive to the couplings between the \(\delta\) meson and nucleons and four-vector meson couplings: smaller couplings lead to a stiffer EOS of NM and thus a larger maximum mass of NSs.
Then, we extended the analysis to three-flavor NS matter with beta equilibrium, where hyperon can appear at the core of NSs, and found that the explicit chiral symmetry breaking term plays an interesting role in determining the mass-radius (M-R) relation of NSs.

The rest of this proceeding is organized as follows.
In Sec. 2, we analyze the properties of NM and the structure of NSs with bELSM based on RMF approximation.
In Sec. 3, we introduce the explicit chiral symmetry breaking term in the bELSM and analyze its effect on the M-R relation of NSs.
Finally, a summary and outlook are given in Sec. 4.

\section{Nuclear properties and neutron star structures}
The Lagrangian, \(\mathcal{L}=\mathcal{L}_{\mathrm{B}}+\mathcal{L}_{\mathrm{M}}+\mathcal{L}_{\mathrm{V}}\), of the bELSM under RMF approximation is given by
\begin{equation}
    \label{eq:lagrangian}
    \begin{aligned}
        \mathcal{L}_{\mathrm{B}}= & \operatorname{Tr}(\bar{B} i \slashed{\partial} B)+c \operatorname{Tr}\left(\bar{B} V \gamma^0 B\right) -g \operatorname{Tr}\left(\bar{B} S^{\prime} B\right)+h \epsilon_{a b c} \epsilon^{d e f} \bar{B}_{a d} \gamma^0 B_{b e} V_{c f} -e \epsilon_{a b c} \epsilon_{d e f} \bar{B}_{a d} B_{b e} S_{c f}^{\prime}\ , \\
        \mathcal{L}_{\mathrm{M}}= & c_2 \operatorname{Tr} S^{\prime 2}-d_2 \operatorname{Tr} \hat{S}^{\prime 2}-c_4 \operatorname{Tr} S^{\prime 4} -2 e_3 \epsilon_{a b c} \epsilon_{d e f} S_{a d}^{\prime} S_{b e}^{\prime} \hat{S}_{c f}^{\prime}\ , \\
        \mathcal{L}_{\mathrm{V}}= & \tilde{h}_2 \operatorname{Tr}\left(S^{\prime 2} V^2\right)+\tilde{g}_3 \operatorname{Tr} V^4 +a_1 \epsilon_{a b c} \epsilon_{d e f} V_{a d} V_{b e}\left(S^{\prime 2}\right)_{c f}\ .
    \end{aligned}
\end{equation}
where \(B\) is the baryon field, \(S^{\prime}\) and \(\hat{S}^{\prime}\) are the 2-quark and 4-quark scalar meson fields, and \(V\) is the vector meson field.
With RMF approximation and 2-flavor assumption, the field representations can be expressed as
\begin{equation}
    S^{\prime}=\frac{1}{\sqrt{2}}\left(\lambda_8 f_0^{\prime}+\lambda_3 a_0^{\prime}\right)+\frac{1}{\sqrt{3}} I \sigma^{\prime}\ ,\quad V=\frac{1}{2}\left(\lambda_3 \rho+\frac{1}{\sqrt{3}} \lambda_8 \omega\right)+\frac{1}{3} I \omega \ ,
\end{equation}
and the 4-quark scalar configuration can be written similarly.
Baryon field can be reduced to \(B \rightarrow \Psi = (p,n)^{\rm T}\).

The spontaneous chiral symmetry breaking can be realized by the vacuum expectation values of the scalar meson fields, which can be obtained by minimizing the effective potential of the scalar meson fields, 
\begin{equation}
    \left\langle\frac{\partial \mathcal{L}_{\mathrm{M}}}{\partial \sigma^{\prime}}\right\rangle=2 \alpha\left(-c_2+2 c_4 \alpha^2+4 e_3 \beta\right)=0\ ,\quad \left\langle\frac{\partial \mathcal{L}_{\mathrm{M}}}{\partial \hat{\sigma}^{\prime}}\right\rangle=2\left(d_2 \beta+2 e_3 \alpha^2\right)=0\ ,
\end{equation}
where \(\langle\sigma^{\prime}\rangle=\sqrt{3} \alpha\) and \(\langle\hat{\sigma}^{\prime}\rangle=\sqrt{3} \beta\).
And the physical scalar states can be obtained by diagonalizing the mass matrix of the scalar meson fields,
\begin{equation}
        \sigma=\cos \theta_0 \sigma^{\prime}+\sin \theta_0 \hat{\sigma}^{\prime}\ ,\quad a_0=\cos \theta_8 a_0^{\prime}+\sin \theta_8 \hat{a_0^{\prime}}\ , \quad f_0=\cos \theta_8 f_0^{\prime}+\sin \theta_8{\hat{f_0}}^{\prime}\ .
\end{equation}

\begin{figure}[htb]
    \centering
    \begin{subfigure}[]{0.3\textwidth}
        \includegraphics[width=\textwidth]{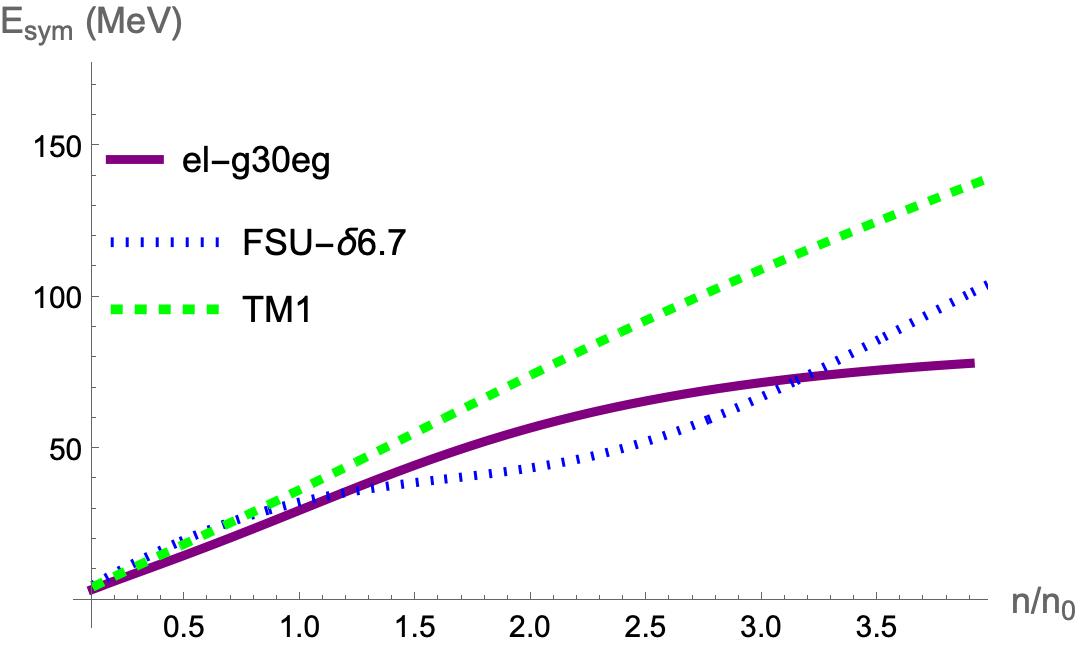}
        \caption{
            Comparison among TM1, FSU-\(\delta 6.7\), and el-g30eg.
        }
        \label{fig:Esymsub1}
    \end{subfigure}
    \begin{subfigure}[]{0.3\textwidth}
        \includegraphics[width=\textwidth]{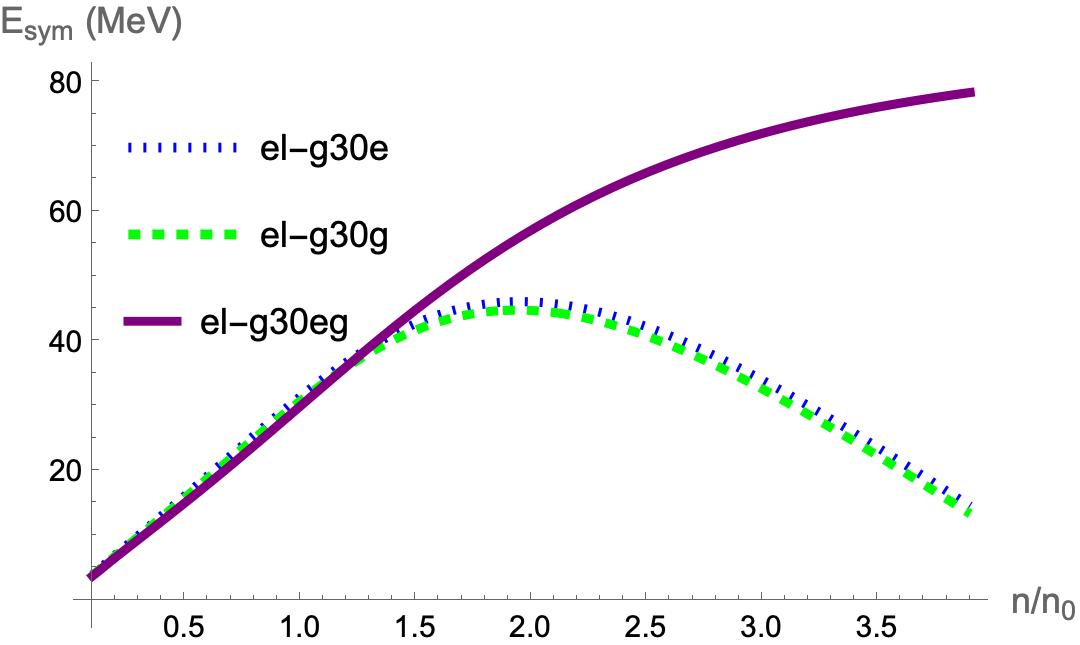}
        \caption{Comparison among cases with different magnitudes of \(g_{a_0 NN}\).}
        \label{fig:Esymsub2}
    \end{subfigure}
    \begin{subfigure}[]{0.3\textwidth}
        \includegraphics[width=\textwidth]{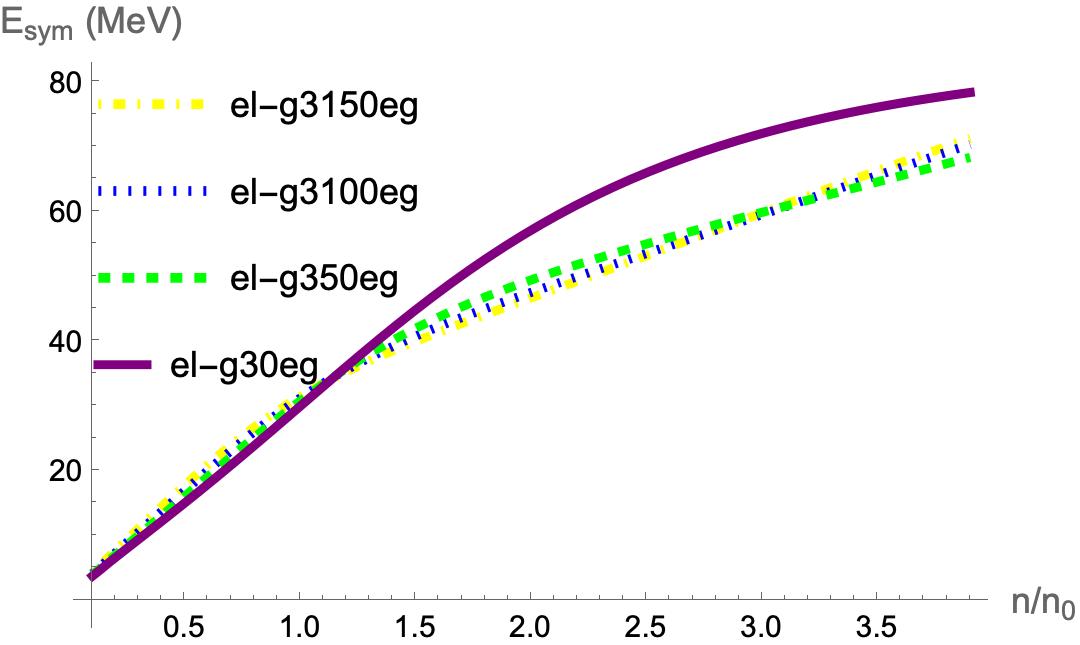}
         \caption{Comparison among cases with different magnitudes of four-vector meson couplings.}
        \label{fig:Esymsub3}
    \end{subfigure}
    \begin{subfigure}[]{0.3\textwidth}
        \includegraphics[width=\textwidth]{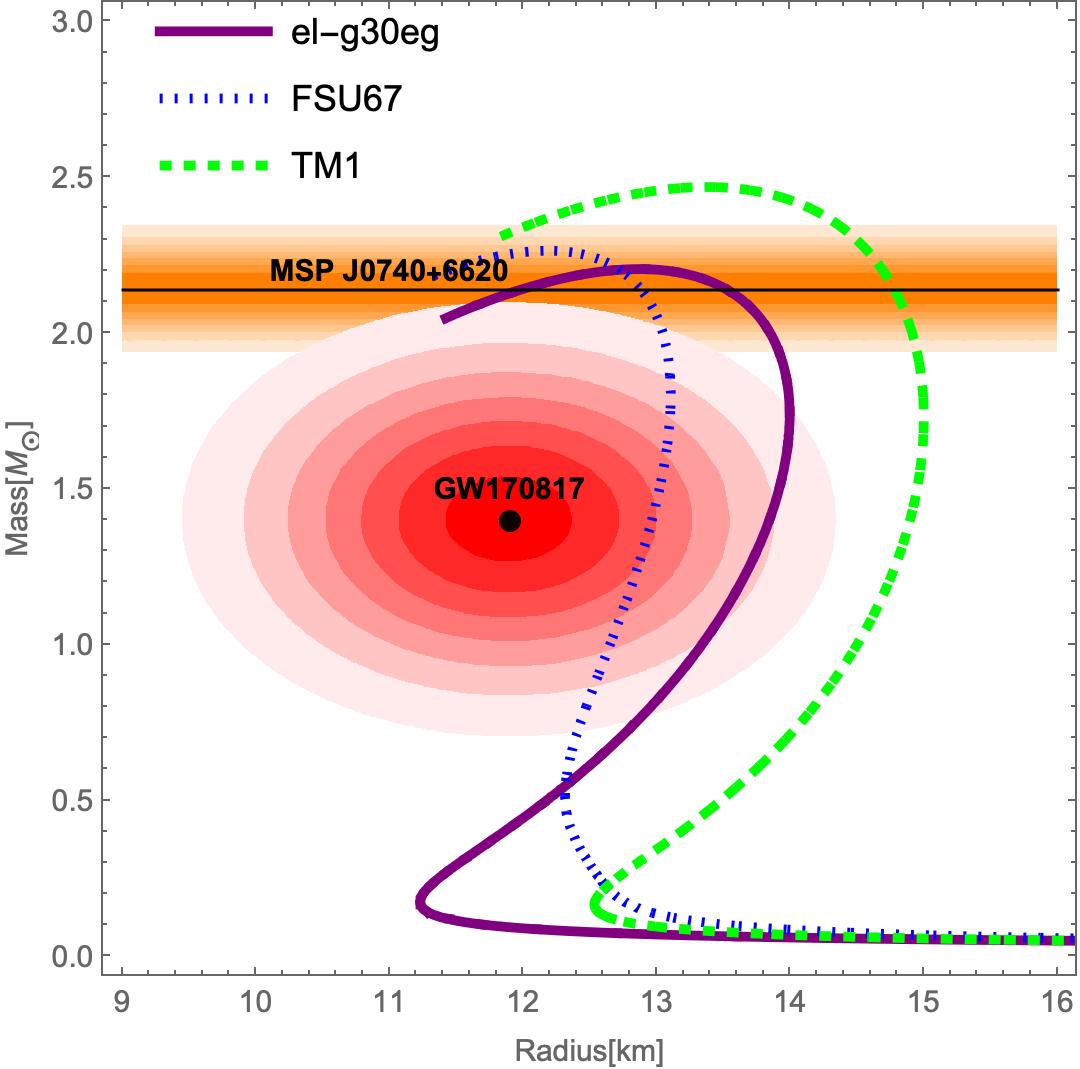}
        \caption{
            Comparison among TM1, FSU-\(\delta 6.7\), and el-g30eg.
        }
        \label{fig:MRsub1}
    \end{subfigure}
    \begin{subfigure}[]{0.3\textwidth}
        \includegraphics[width=\textwidth]{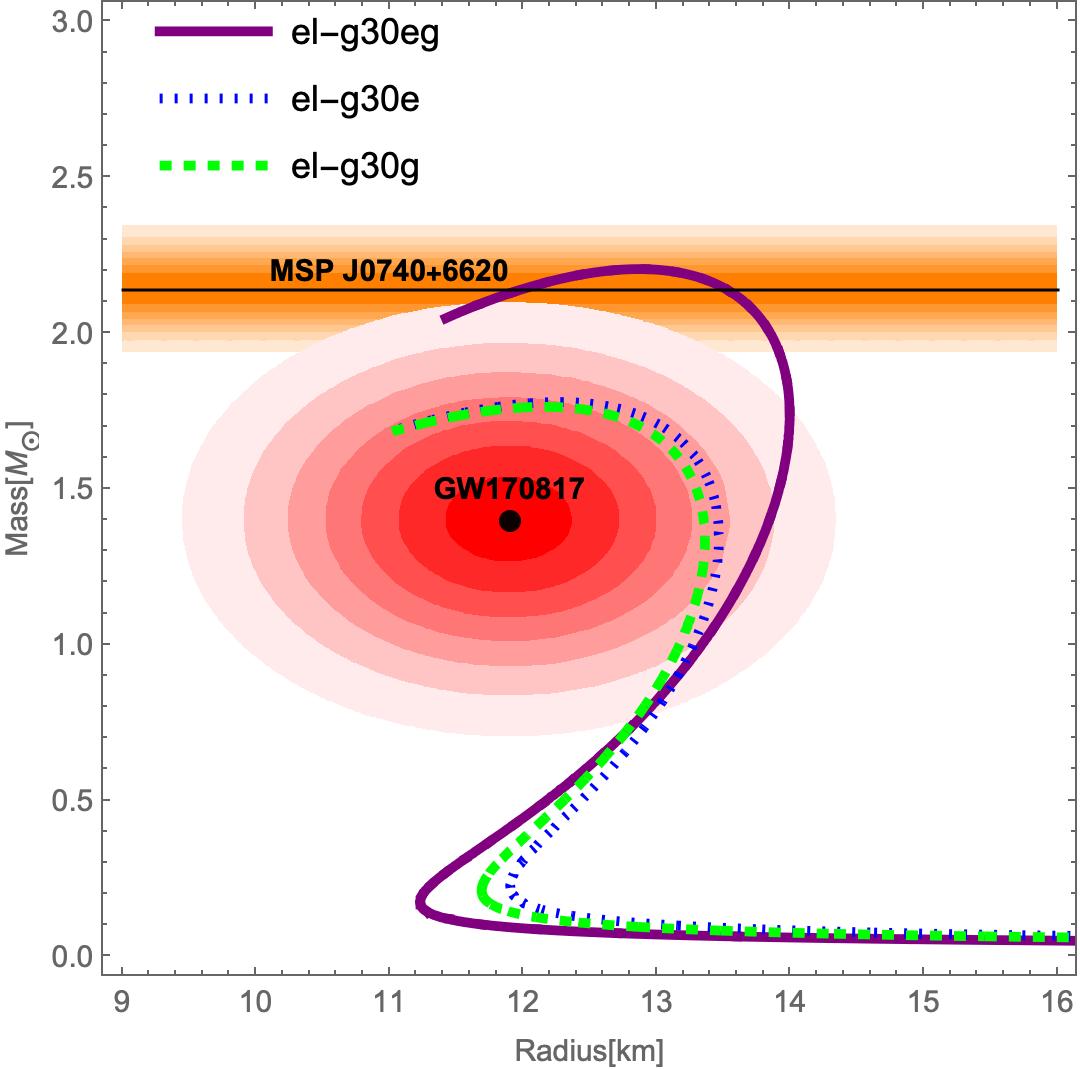}
        \caption{Comparison among cases with different magnitudes of \(g_{a_0 NN}\).}
        \label{fig:MRsub2}
    \end{subfigure}
    \begin{subfigure}[]{0.3\textwidth}
        \includegraphics[width=\textwidth]{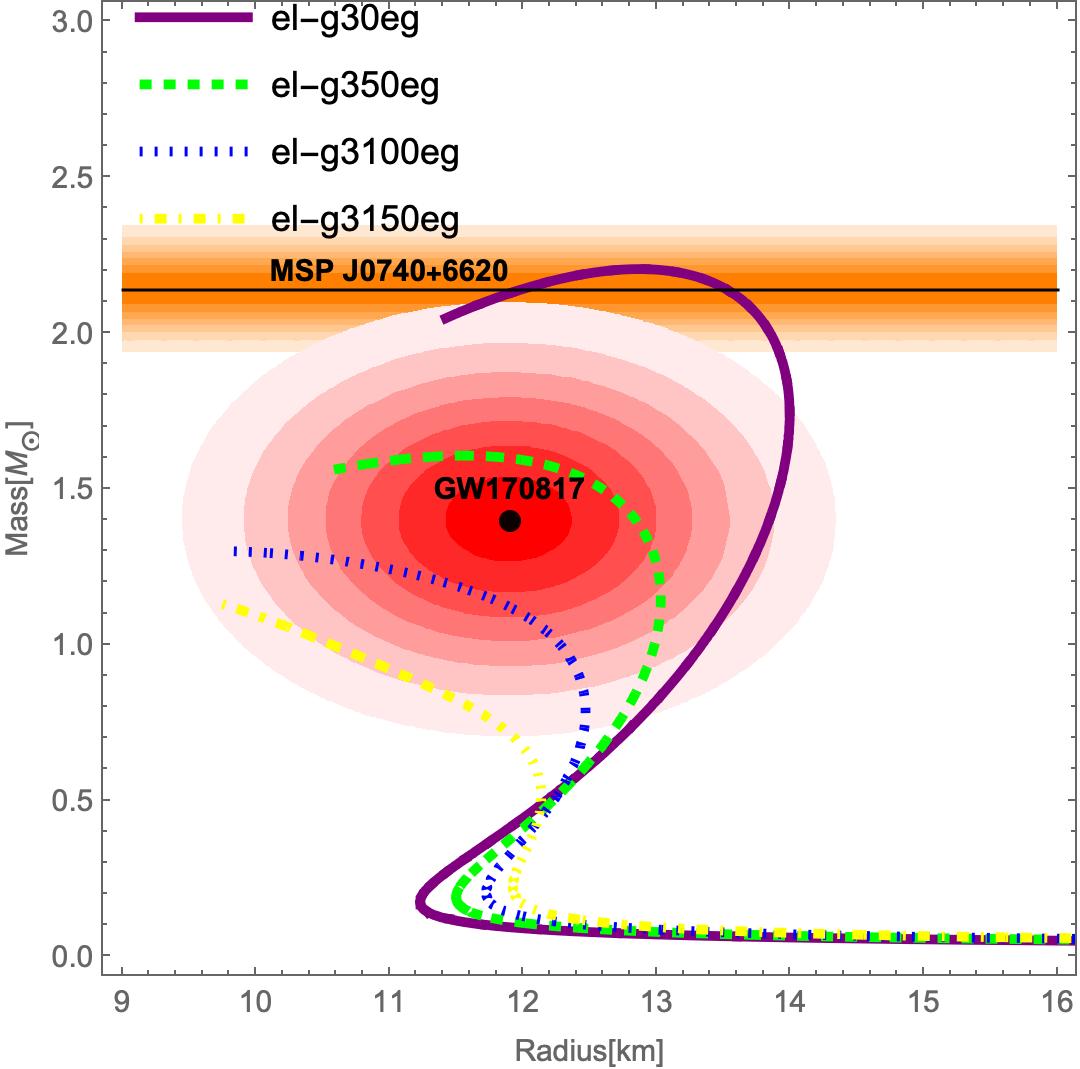}
        \caption{Comparison among cases with different magnitudes of four-vector meson couplings.}
        \label{fig:MRsub3}
    \end{subfigure}
    \caption{
        The \(E_{\rm sym}(n)\) and M-R relation of NSs for different cases.
        The constraints MSP J0740+6620 are from Ref.~\cite{NANOGrav:2019jur} and GW170817 are from Ref.~\cite{LIGOScientific:2018cki}.
        Both are at \(95\%\) confidence level.
    }
    \label{fig:2-flavor}
\end{figure}

After fitting the parameters of the model to the spectra of hadrons at vacuum and the properties of NM around saturation density, see details in Ref.~\cite{Ma:2025llw}, following parameter sets are chosen for the analysis, shown in Table \ref{tab:g}, and compared to the results of TM1~\cite{Sugahara:1993wz} and FSU-\(\delta\)6.7~\cite{Li:2022okx}.
\begin{table}[htb]\small
	\caption{
        The values of meson-baryon coupling in various of models.
        The "el-" prefix indicates the extended linear sigma model.
    }
    \label{tab:g}
    \centering
	\begin{threeparttable}
		\begin{tabular}{@{}cccccccc}
			\hline
			\hline
			& $g_{\sigma NN}$ & \(g_{{a_0 NN}}\) & $g_{fNN}$ & $g_{\omega NN}$ & $g_{\rho NN}$ & $\tilde{g}_3$ \\
			\hline
			TM1 & $-10.0$ & ---& --- & $-12.6$ & $-4.63$ & $23.8$ \\
            \hline
			FSU-\(\delta\)6.7 & $10.2$ & $6.7$ & --- & $-13.4$ & $-7.27$ & $43.0$ \\
			\hline
			el-g30g & $-6.17$ & $5.13$ & $2.95$ & $-6.09$ & $5.30$ & $1.59$ \\
            \hline
			el-g30e & $-6.20$ & $-5.03$ & $3.00$ & $6.09$ & $5.30$ & $0.542$ \\
            \hline
			el-g30eg & $-5.97$ & $-0.671$ & $2.68$ & $6.06$ & $3.45$ & $0.397$ \\
            \hline
			el-g350eg & $-6.12$ & $-0.852$ & $2.85$ & $6.37$ & $3.71$ & $51.5$ \\
            \hline
			el-g3100eg & $-6.36$ & $-0.442$ & $3.19$ & $6.73$ & $3.85$ & $100$ \\
            \hline
			el-g3150eg & $-6.38$ & $-0.413$ & $3.20$ & $7.09$ & $4.04$ & $150$ \\
			\hline
			\hline
		\end{tabular}
	\end{threeparttable}
\end{table}
Then, the resulting \(E_{\rm sym}(n)\) and M-R relation of NSs are shown in Fig. \ref{fig:2-flavor}.
We have the following findings from the results:
\begin{itemize}
    \item From Figs.~\ref{fig:Esymsub1} and~\ref{fig:MRsub1}, we can see that the introduction of the \(\delta\) meson makes the \(E_{\rm sym}(n)\) softer at intermediate density regions and thus leads to a smaller radius of the \(1.4M_{\odot}\) NS, which is favored by the astrophysical observations and align with the results of FSU-\(\delta\)6.7.
    \item From Figs.~\ref{fig:Esymsub2} and~\ref{fig:MRsub2} and Table~\ref{tab:g}, we can see that the coupling between the \(\delta\) meson and nucleons, \(g_{a_0 NN}\), has a significant effect on the maximum mass of NSs, that a smaller \(g_{a_0 NN}\) leads a large enough maximum mass of NSs to satisfy the constraint from MSP J0740+6620 around \(2M_{\odot}\).
    \item From Figs.~\ref{fig:Esymsub3} and~\ref{fig:MRsub3} and Table~\ref{tab:g}, we can see that the four-vector meson couplings, \(\tilde{g}_3\) also affects the maximum mass of NSs, that the smaller couplings results into a larger maximum mass of NSs, but incompressibility at saturation density, \(K_0\), is around \(500\) MeV, larger than the empirical value, \(200-300\) MeV~\cite{Sedrakian:2022ata}.
\end{itemize}

\section{Explicit chiral symmetry breaking and neutron star structures}
\begin{figure}[htpb]
    \centering
        \includegraphics[width=0.35\textwidth]{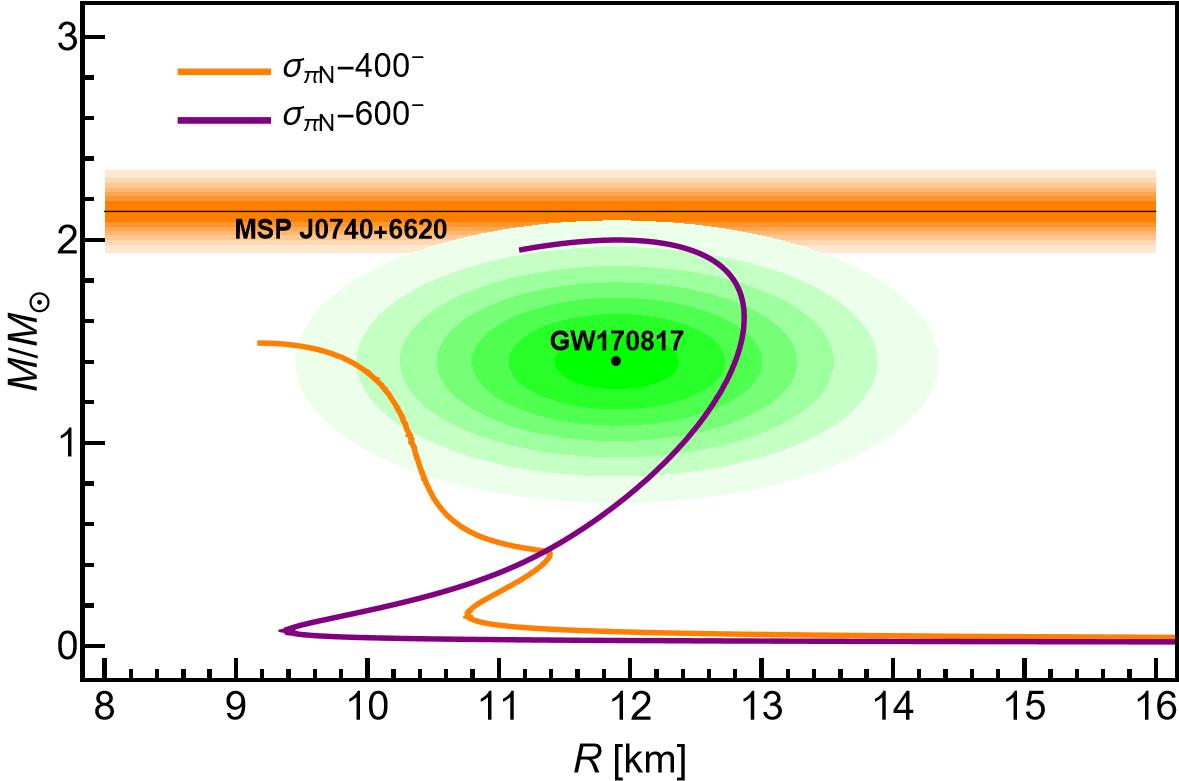}
        \includegraphics[width=0.35\textwidth]{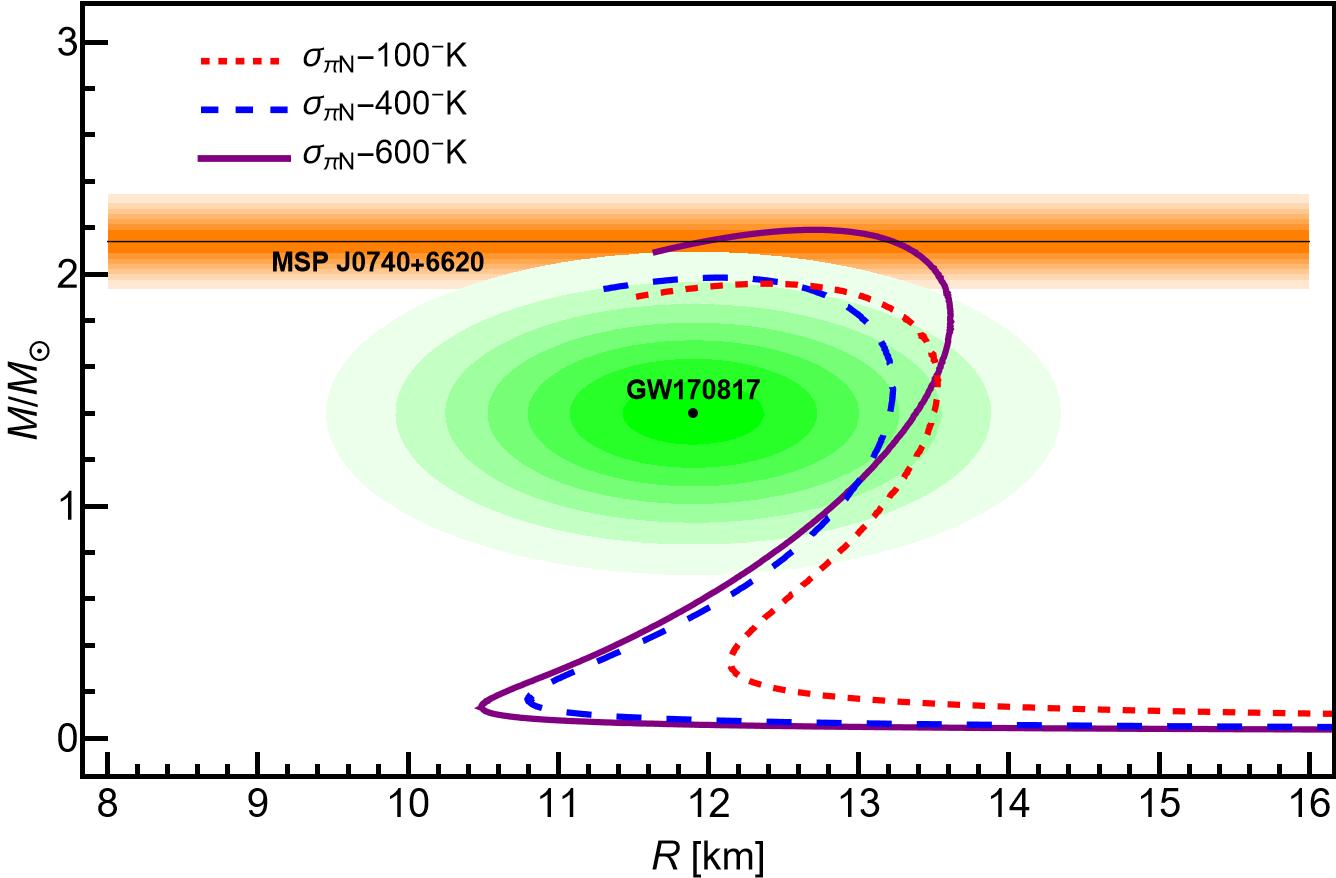}
    \caption{
        The NS structures with \(\sigma_{\pi N}\)-\(100^-(K)\), \(\sigma_{\pi N}\)-\(400^-(K)\) and \(\sigma_{\pi N}\)-\(600^-(K)\) parameter sets, where the number is \(-\sigma_{\pi N}\) in MeV and \(K\) denotes \(K_0\approx 500\) MeV.
        \(K_0=521.9,\,520.3,\,518.8\) MeV respectively; \(K_0\approx 240\) MeV for non-\((K)\) sets.
        \(\sigma_{\pi N}=m_N-m\) is the pion-nucleon sigma term. Constraints as in Fig.~\ref{fig:2-flavor}.
    }
    \label{fig:3-flavor}
\end{figure}

After introducing the explicit chiral symmetry breaking term in the bELSM with a constant background field, \(\xi\), and neglecting the 4-quark configuration for simplicity, the Lagraingian becomes
\begin{equation}
    \begin{aligned}
        \mathcal{L}_{\mathrm{M}}= & c_2 \operatorname{Tr} S^{\prime 2}-c_4 \operatorname{Tr} S^{\prime 4} -b_1 \operatorname{Tr}\left(\xi S^{\prime 3}\right)-2~G \operatorname{Tr}\left(\xi S^{\prime}\right)\ , \\
        \mathcal{L}_{\mathrm{V}}= & \tilde{h}_2 \operatorname{Tr}\left(V^2 S^{\prime 2}\right)+a_1 \epsilon^{i j k} \epsilon^{\operatorname{lmn}}(V)_{i l}(V)_{j m}\left(S^{\prime 2}\right)_{k n} +b_2 \operatorname{Tr}\left(V^2 \xi S^{\prime}\right)+b_3 \epsilon^{i j k} \epsilon^{\operatorname{lmn}}(V)_{i l}(V)_{j m}\left(\xi S^{\prime}\right)_{k n} \\
        & +\tilde{g}_3 \operatorname{Tr} V^4+\tilde{a}_2 \epsilon^{i j k} \epsilon^{\operatorname{lmn}}(V)_{i l}(V)_{j m}\left(V^2\right)_{k n}\ , \\
        \mathcal{L}_{\mathrm{B}}= & \operatorname{Tr}\left(\bar{B} i \gamma_\mu \partial^\mu B\right)+c \operatorname{Tr}\left[\bar{B} \gamma_0 V B\right]+c^{\prime} \operatorname{Tr}\left[\bar{B} \gamma_0 B V\right] +h \epsilon^{i j k} \epsilon^{\operatorname{lmn}}(\bar{B})_{i l}\left(\gamma_0 V\right)_{j m}(B)_{k n}-g \operatorname{Tr}\left[\bar{B} S^{\prime} B\right] \\
        & -e \epsilon^{i j k} \epsilon^{\operatorname{lmn} n}(\bar{B})_{i l}\left(S^{\prime}\right)_{j m}(B)_{k n}-b_4 \operatorname{Tr}[\bar{B} \xi B] -b_5 \epsilon^{i j k} \epsilon^{\operatorname{lmn}}(\bar{B})_{i l}(\xi)_{j m}(B)_{k n}\ ,
    \end{aligned}
\end{equation}
where \(\xi=\frac{1}{\sqrt{2}}(\lambda_8 \xi_8+\lambda_3 \xi_3)+\frac{1}{\sqrt{3}} I \xi_0\), which makes the symmetry breaking pattern follow as
\begin{equation}
    S^{\prime}=\frac{1}{\sqrt{2}}\left[\left(\alpha_3+a_0\right) \lambda_3+\left(\alpha_8+f_0\right) \lambda_8\right]+\frac{1}{\sqrt{3}} I\left(\alpha_0+\sigma\right)\ ,
\end{equation}
where \(\alpha_0\), \(\alpha_3\), and \(\alpha_8\) are the vacuum expectation values of the scalar meson fields.
By fitting the parameters to the spectra of relevant baryons and mesons at vacuum, see details in Ref.~\cite{Ma:2025tyt}, the resulting M-R relation of NSs is shown in Fig.~\ref{fig:3-flavor}.
We have the following findings from the results:
\begin{itemize}
    \item A larger explicit chiral symmetry breaking term can lead to a stiffer EOS of NS matter and thus a larger maximum mass of NSs, and the \(\sigma_{\pi N}\sim -600~\rm MeV\), which deviates from the empirical value, \((32-89)~\rm MeV\)~\cite{Bernard:1995dp,Meissner:2005ba}, is needed to satisfy the constraint from MSP J0740+6620 around \(2M_{\odot}\).
    \item A larger incompressibility at saturation density, \(K_0\), can also lead to more observation-favored NS M-R relations.
\end{itemize}
    
\section{Summary and outlook}
The properties of NM and the structure of NSs are analyzed with the bELSM under RMF approximation, where mass spectra of hadrons are obtained via the chiral symmetry pattern of the low-energy strong interaction.
The plateau structure of \(E_{\rm sym}(n)\) at intermediate densities is found with the introduction of the \(\delta\) meson, which is crucial to the consistency of neutron skin thickness of \(^{208}\)Pb and the tidal deformability of a canonical NS, aligning with the previous studies.
The mass maximum mass of NSs is found to be sensitive to the \(g_{a_0 NN}\) and four-vector meson couplings, and both should deviate from the empirical values, at vacuum and saturation density, to satisfy the astrophysical constraints.
Then, the explicit chiral symmetry breaking term is introduced in the bELSM to discuss the beta-equilibrium NS matter, and it's found that by tuning \(\sigma_{\pi N}\) value, a stiffer EOS of NS matter can be achived, leading to a larger maximum mass of NSs, but the value of \(\sigma_{\pi N}\) needed to satisfy the constraint from MSP J0740+6620 around \(2M_{\odot}\) is negative, not positive as the vacuum value.
All the findings indicate that if we want to reproduce the NS structures with the consideration of the connection to the QCD symmetries, the running behaviors of the parameters should be considered to give the density dependence of the low-energy theory/model.

Fortunately, the studies in this proceeding offers the possible running behavior of the corresponding parameters, and may be a good starting point for the future studies.
In the future, we plan to include the density dependence of the parameters in the bELSM by including the quantum corrections to RMF approximation to analyze the properties of NM and the structure of NSs.

\section*{Acknowledgments}
The work of Y. M. is supported by Jiangsu Funding Program for Excellent Postdoctoral Talent under Grant Number 2025ZB516.

\bibliography{baryon}
\bibliographystyle{cas-model2-names}

\end{document}